\documentclass[prl,twocolumn,showpacs,groupedaddress]{revtex4}
\usepackage[dvips]{epsfig}
\usepackage{subfigure}
\usepackage{float}
\usepackage{amsmath}
\usepackage{amssymb}
\usepackage{amsfonts}
\usepackage{rotating}
\usepackage{euscript}
\usepackage{subfigure}
\usepackage{enumerate}
\usepackage{hhline}
\usepackage{threeparttable}
\usepackage{supertabular}
\usepackage{pslatex}
\usepackage{tabularx}

\begin{document}

\bibliographystyle{apsrev}

\title{\bf{Optical loss and lasing characteristics of high-quality-factor AlGaAs microdisk resonators with embedded quantum dots}}

\author{Kartik Srinivasan}
\email{phone: (626) 395-6269, fax: (626) 795-7258, e-mail: kartik@caltech.edu}
\author{Matthew Borselli}
\author{Thomas J. Johnson}
\author{Paul E. Barclay}
\author{Oskar Painter}
\affiliation{Thomas J. Watson, Sr. Laboratory of Applied Physics, California Institute of Technology, Pasadena, CA 91125, USA.}
\author{Andreas Stintz}
\author{Sanjay Krishna}
\affiliation{Center for High Technology Materials, University of New Mexico, Albuquerque, NM 87106, USA.}

\date{\today}

\begin{abstract}  
\noindent Optical characterization of AlGaAs microdisk resonant cavities with a quantum dot active region is presented.  Direct passive measurement of the optical loss within AlGaAs microdisk resonant structures embedded with InAs/InGaAs dots-in-a-well (DWELL) is performed using an optical-fiber-based probing technique at a wavelength ($\lambda\sim1.4$ $\mu$m) that is red-detuned from the dot emission wavelength ($\lambda\sim1.2$ $\mu$m).  Measurements in the $1.4$ $\mu$m wavelength band on microdisks of diameter $D=4.5$ $\mu$m show that these structures support modes with cold-cavity quality factors as high as $3.6{\times}10^5$.  DWELL-containing microdisks are then studied through optical pumping at room temperature.  Pulsed lasing at $\lambda\sim1.2$ $\mu$m is seen for cavities containing a single layer of InAs dots, with threshold values of $\sim17$ $\mu$W, approaching the estimated material transparency level.  Room-temperature continuous wave operation is also observed.  
\end{abstract}

\pacs{42.70.Qs, 42.55.Sa, 42.60.Da, 42.55.Px}
\maketitle

Recently, multiple research groups have demonstrated vacuum Rabi splitting in a semiconductor system consisting of a single quantum dot (qdot) exciton embedded in an optical microcavity\cite{ref:Reithmaier,ref:Yoshie3,ref:Peter}. These experiments have in many ways confirmed the potential of semiconductor microcavities for chip-based cavity quantum electrodynamics (cQED) experiments.  For future experiments, such as those involving quantum state transfer in quantum networks\cite{ref:Cirac}, it will be important to further improve upon the parameters of such qdot-microcavity systems.  One clear improvement required is to move the system further within the regime of strong coupling\cite{ref:Kimble2}, defined as having the atom-photon coupling rate ($g$) larger than both the cavity loss rate ($\kappa$) and qdot dipole decay rate ($\gamma$).  In particular, the ratio of $g$ to the larger of $\kappa$ and $\gamma$ approximately represents the number of Rabi oscillations that can take place before the effects of dissipation destroy coherent energy exchange\cite{ref:Kimble2}.  In each of Refs. \cite{ref:Reithmaier,ref:Yoshie3,ref:Peter}, loss in the system was found to be dominated by the optical cavity, with $g \lesssim \kappa$.  As the low-temperature homogeneous linewidth in self-assembled InAs qdots is typically a few $\mu$eV\cite{ref:Bayer}, corresponding to a qdot dipole decay rate of $\gamma/2\pi \sim 1$ GHz, it will be advantageous to develop cavities with quality factors such that $\kappa/2\pi \lesssim 1$ GHz, with further improvements in $Q$ serving mainly to improve the optical collection efficiency of emitted light. For the $\lambda\sim 0.9$-$1.2$ $\mu$m emission wavelength for InAs qdots\cite{ref:Reithmaier,ref:Yoshie3}, this corresponds to an optical mode quality factor of $Q \sim 1 {\times}10^5$ ($\kappa/2\pi=\omega/4{\pi}Q$).  Achieving such low loss cavities is also important in light of the difficulty in fabricating a structure where the qdot is \emph{optimally} positioned for maximum coupling to the cavity mode.

\begin{figure}
\begin{center}
\epsfig{figure=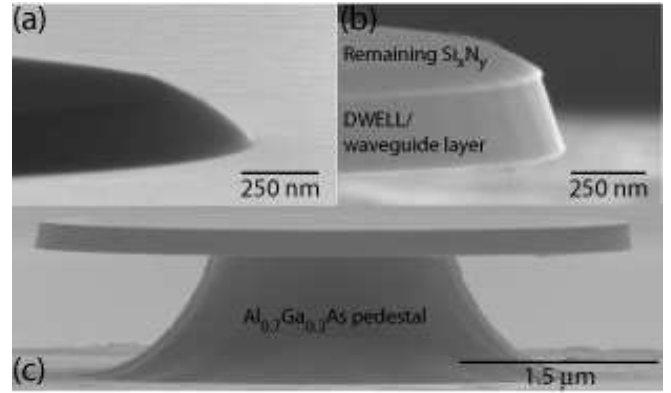, width=\linewidth}
\caption{Scanning electron microscope (SEM) images of DWELL-containing microdisk cavities after the (a) Si$_{x}$N$_{y}$ etch, and (b)-(c) AlGaAs etch and undercut.} 
\label{fig:disk_SEMs}
\end{center}
\end{figure}

Alongside the work on cQED in qdot-microcavity systems, there has been significant progress in developing higher $Q$ and smaller effective mode volume ($V_{\text{eff}}$) semiconductor microcavities over the last few years.  A variety of geometries and materials have been studied, ranging from InP photonic crystal microcavities ($Q\sim1.3{\times}10^4$, $V_{\text{eff}}\sim(\lambda/n)^3$)\cite{ref:Srinivasan4} to Si photonic crystal\cite{ref:Akahane2,ref:Srinivasan7} ($Q\sim4{\times}10^4$, $V_{\text{eff}}\sim(\lambda/n)^3$) and microdisk\cite{ref:Borselli} ($Q\sim5{\times}10^5$,$V_{\text{eff}}\sim5(\lambda/n)^3$) cavities.  Of particular importance to the self-assembled InAs qdot work is the host AlGaAs material system.  Here, we report the creation of $D$=4.5 $\mu$m diameter AlGaAs microdisks that exhibit $Q$ factors as high as $3.6{\times}10^5$ at $\lambda\sim1.4$ $\mu$m, a value which, to our knowledge, exceeds the highest $Q$ factors measured for AlGaAs microcavities to date\cite{ref:Gayral,ref:Michler2,ref:Reithmaier,ref:Yoshie3}.  These AlGaAs microdisks contain embedded quantum dots-in-a-well (DWELL)\cite{ref:Krishna,ref:Liu_G} which have a ground-state emission at $\lambda\sim1.2$ $\mu$m, so the passive measurements at $\lambda\sim1.4$ $\mu$m are performed where the qdots are relatively non-absorbing. The lasing characteristics of these devices are also investigated through photoluminescence measurements, and low threshold, room temperature qdot lasers are demonstrated.      

\begin{figure}
\begin{center}
\epsfig{figure=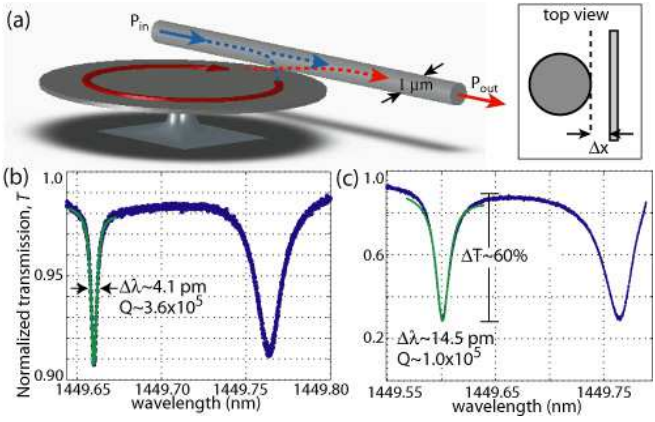, width=\linewidth}
\caption{(a) Schematic geometry for probing the microdisk cavities through side-coupling via an optical fiber taper. (b) Normalized taper transmission ($T=P_{out}/P_{in}$) of a $4.5$ $\mu$m diameter microdisk for a taper-disk lateral separation (${\Delta}x$) of $\sim800$ nm.  (c) Normalized taper transmission for the same device with ${\Delta}x\sim200$ nm. The solid green curves are Lorentzian fits to the data.}
\label{fig:passive_Q}
\end{center}
\end{figure}


The microdisks studied here are fabricated from an epitaxy consisting of one or three layers of InAs qdots embedded in (1-3) In$_{0.15}$Ga$_{0.85}$As quantum wells, which are in turn sandwiched between GaAs/Al$_{0.3}$Ga$_{0.7}$As layers to create a total waveguide layer that is $d$=$255$ nm thick.  This layer is grown on top of a $1.5$ $\mu$m Al$_{0.7}$Ga$_{0.3}$As sacrificial layer that forms the supporting pedestal (Fig. \ref{fig:disk_SEMs}(c)).  The cavities are created through: (i) electron beam lithography and subsequent reflow of the e-beam resist to produce smooth and circular patterns, (ii) SF$_6$/C$_4$F$_8$ inductively-coupled plasma reactive ion etching (ICP-RIE) of a deposited Si$_{x}$N$_{y}$ mask layer (Fig. \ref{fig:disk_SEMs}(a)), (iii) Ar-Cl$_2$ ICP-RIE etching of the Al$_{0.3}$Ga$_{0.7}$As layer and removal of the remaining Si$_{x}$N$_{y}$ layer, and (iv) wet chemical etching of the underlying Al$_{0.7}$Ga$_{0.3}$As layer to form the supporting pedestal (Fig. \ref{fig:disk_SEMs}(b)-(c)). The Si$_{x}$N$_{y}$ etch step is particularly important, as any roughness in this mask layer is transferred into underlying layers.  This etch has thus been calibrated to produce as smooth a sidewall surface as possible (Fig. \ref{fig:disk_SEMs}(a)), without particular concern for its verticality. The subsequent Ar-Cl$_2$ etch is highly selective so that the angled mask does not result in erosion of the AlGaAs sidewalls.  

Initial passive measurements to measure the cold-cavity $Q$ factor of the microdisk resonant modes were performed using an optical-fiber-based evanescent coupling technique\cite{ref:Srinivasan7,ref:Borselli,ref:Kippenberg}.  An optical fiber taper is formed by heating and adiabatically stretching a standard single mode fiber until it reaches a minimum diameter $\sim 1$ $\mu$m.  A fiber-coupled scanning tunable laser ($<5$ MHz linewidth) is spliced to the taper's input, and when the taper is brought within a few hundred nanometer (nm) of the cavity, their evanescent fields interact, and power transfer can result.  A schematic illustrating the coupling geometry for this system is shown in Fig. \ref{fig:passive_Q}(a).  Further details of the mounting and positioning of the fiber taper are described in Refs. \cite{ref:Srinivasan7,ref:Borselli}.  The $Q$ of a cavity mode is determined by examining the linewidth of the resulting resonance in the taper's wavelength dependent transmission spectrum. In Fig. \ref{fig:passive_Q}(b), we show a ``doublet'' resonance of a microdisk ($D$=4.5 $\mu$m, 1-DWELL epitaxy) in the $1400$ nm wavelength band when the taper is $\sim 1$ $\mu$m to the side of the disk; the separation is kept large in order to reduce taper loading effects\cite{ref:Srinivasan7,ref:Borselli}. The double resonance peaks correspond to standing wave modes formed from mixtures of the degenerate clockwise and counterclockwise whispering gallery modes that couple and split due to the disk-edge surface roughness\cite{ref:Borselli,ref:Kippenberg,ref:Weiss}. The linewidth ($\Delta\lambda$) of the shorter wavelength resonance corresponds to $Q\sim3.6{\times}10^5$.  Similarly, in Fig. \ref{fig:passive_Q}(c), we show the spectral response of the doublet when the taper is positioned much closer ($\sim$ 200 nm) to the edge of the disk, so that the amount of coupling has increased.  The combination of increased coupling as well as parasitic loading due to the presence of the taper has increased the total loss rate of the resonant mode, yielding a loaded $Q\sim1.0{\times}10^5$.  The depth of coupling, however, has also considerably increased from $10\%$ to $60\%$, corresponding to a photon collection efficiency (the ratio of ``good'' coupling to all other cavity losses including parasitic and intrinsic modal loss) of approximately $20\%$.  It is believed that the high $Q$ values achieved in these measurements are due to a combination of the resist reflow process that reduces radial variations and subsequent Rayleigh scattering in the disk\cite{ref:Borselli}, and the optimized dry etching processes that create very smooth disk-edge sidewalls.        

The demonstrated $Q$ is high enough that, if used in cQED, the cavity will have a decay rate $\kappa/2\pi\sim0.35$ GHz (at $\lambda \sim 1.2$ $\mu$m), lower than the aforementioned typical low temperature qdot dipole decay rate of $\gamma/2\pi\sim1$ GHz.  After adjusting for the reduced wavelength of the qdot resonance, the current devices ($D$=4.5 $\mu$m) have a $V_{\text{eff}} \sim 6(\lambda/n)^3$ for the standing wave resonant modes studied here\footnote{Note that $V_{\text{eff}}$ for a standing wave whispering gallery mode is half that of a traveling wave mode.}.  For a maximally coupled InAs qdot (spontaneous emission lifetime $\tau\sim1$ ns, oscillator strength $f\sim18$\cite{ref:Andreani}), this mode volume corresponds to $g/2\pi\sim11$ GHz.  Thus, even for the disk sizes considered in this work, an appropriately positioned qdot would place the system deep within the strong coupling regime.  Of additional importance is the fiber-based coupling technique used here.  This method allows for the $Q$ to be accurately determined in a way that does not rely upon the (weak) background emission from the qdots\cite{ref:Reithmaier,ref:Yoshie3,ref:Peter}; all that is required is a probe laser that can be slightly detuned from the qdot absorption lines.  Furthermore, the taper also acts as a coupler that transfers light from an optical fiber into the wavelength-scale mode volume of the cavity, where it can interact with the qdots, and as a subsequent output coupler.  Such integration could markedly improve the collection efficiency in cQED experiments, particularly important for microdisk and photonic crystal cavities, which typically do not have a radiation pattern that can be effectively collected by free-space optics or a cleaved fiber\cite{ref:Yoshie3}.

\begin{figure}
\begin{center}
\epsfig{figure=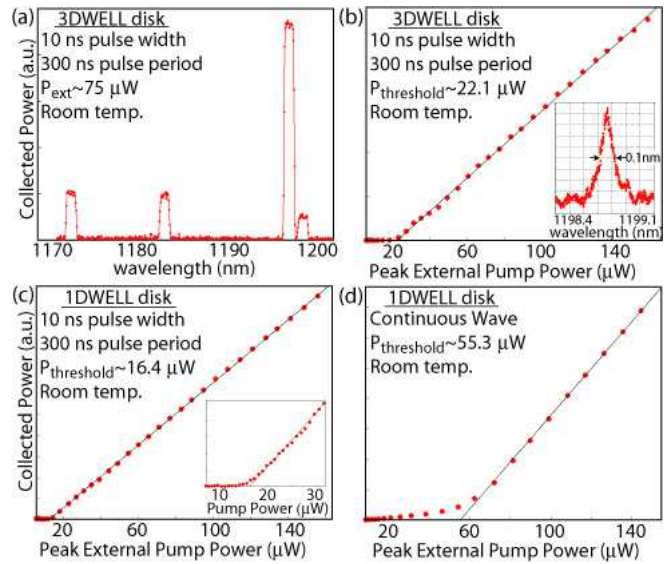, width=\linewidth}
\caption{(a) Photoluminescence spectrum of a 3-DWELL microdisk device (OSA resolution bandwidth (RBW)$=1$ nm). (b)-(d) L-L curves for: (b) pulsed 3-DWELL microdisk laser (inset shows the sub-threshold spectrum of cavity mode with a resolution-limited (RBW=0.10 nm) linewidth), (c) pulsed 1-DWELL microdisk laser (inset shows L-L curve near threshold), and (d) 1-DWELL microdisk laser under CW pumping conditions. The dashed lines are least-square linear fits to the above-threshold data.}
\label{fig:laser_measurements}
\end{center}
\end{figure}

In addition to the fiber-based passive measurements of the microdisks at $\lambda\sim1.4$ $\mu$m, we performed room temperature photoluminescence measurements to study the qdot emission in the $1.2$ $\mu$m wavelength band.  The cavities ($D$=5 $\mu$m in this case) were optically pumped at room temperature using a pulsed 830 nm semiconductor laser, and the emitted laser light was collected by a microscope objective and spectrally resolved in an optical spectrum analyzer (OSA).  Initial measurements were performed on cavities containing 3 DWELLs due to their higher modal gain, roughly three times that of a single DWELL layer\cite{ref:Liu_G}.  Emission is observed for a few ($\sim$2-5) modes in a given microdisk (Fig. \ref{fig:laser_measurements}(a)), and the linewidths of the resonant modes (taken at sub-threshold pump powers) are as narrow as the resolution limit of the OSA (inset of Fig. \ref{fig:laser_measurements}(b)).  Fig. \ref{fig:laser_measurements}(b) shows a typical light-in-light-out (L-L) curve for a 3-DWELL device pumped with a 300 ns period and 10 ns pulse width; the device exhibits lasing action with an estimated threshold value of $\sim$22 $\mu$W.

The saturated ground state modal gain for single DWELL epitaxies has been estimated to be $\sim3.6$-$5.4$ cm$^{-1}$\cite{ref:Liu_G,ref:Eliseev}.  Noting that modal gain approximately equals modal loss at threshold, this indicates that a \emph{minimum} cavity $Q \sim 3$-$5 \times 10^4$ is required for this single layer of qdots to provide enough gain compensation to achieve lasing.  The fiber-based linewidth measurements described earlier indicate that such $Q$ factors should be achievable, and indeed, lasing from the qdot ground states is observed in these single dot layer devices (Fig. \ref{fig:laser_measurements}(c)).  The laser threshold pump power for the 1-DWELL devices was measured to be as small as $16.4$ $\mu$W, significantly lower than the $750$ $\mu$W threshold values recently reported for similarly sized microdisk qdot lasers\cite{ref:Ide}.  Furthermore, as shown in Fig. \ref{fig:laser_measurements}(d), continuous wave (CW), room temperature lasing was also obtained, albeit with a somewhat higher laser threshold.   

The laser threshold values we report here are the peak pump powers incident on the sample surface; the absorbed power is estimated to be roughly 16$\%$ of this value, determined by calculating the expected reflectivities at the disk interfaces and assuming an absorption coefficient of 10$^4$ cm$^{-1}$ in the GaAs and quantum well layers\cite{ref:Coldren}. The threshold absorbed pump power for the 1-DWELL lasers is thus $\sim 2.6$ $\mu$W.  From this, the equivalent threshold current density, useful for comparing the performance of the microdisk lasers to previously demonstrated broad-area stripe lasers, can be estimated.  Given the pump spot size ($\sim16$ $\mu$m$^{2}$), and assuming an internal quantum efficiency $\sim1$, we arrive at an equivalent threshold current density of $11$ A/cm$^2$ for the 1-DWELL devices.  In comparison, the estimated \emph{transparency} current density in previous work on broad-area 1-DWELL lasers was 10.1 A/cm$^2$\cite{ref:Liu_G}.  The proximity of the demonstrated laser threshold to this transparency value indicates that non-intrinsic optical losses within the microdisk cavity have largely been eliminated.    

In conclusion, AlGaAs microdisks as small as $4.5$ $\mu$m in diameter and supporting standing wave resonant modes with $Q$ factors as high as $3.6{\times}10^5$ in the $1400$ nm wavelength band have been demonstrated.  These cavities contain integral InAs quantum dots, and initial room temperature photoluminescence measurements have yielded laser threshold values as low as $16.4$ $\mu$W, nearing the transparency level of the material.  Future efforts will be directed towards low-temperature, near-resonance studies of the quantum dot-microdisk system and further studies of the trade-off between microdisk size and optical loss.  In addition, use of the fiber taper coupler will be examined as a tool that can greatly facilitate such experiments.

K.S. thanks the Hertz Foundation and M.B. thanks the Moore Foundation, NPSC, and HRL Laboratories for their graduate fellowship support. 
\bibliography{./PBG}

\end{document}